\documentclass[]{spie}  

\usepackage{amsmath,amsfonts,amssymb}
\usepackage{graphicx}
\usepackage[colorlinks=true, allcolors=blue]{hyperref}

\title{INGOT WFS for LGSs: First Results from Simulations}

\author[a,b]{Elisa Portaluri}
\author[a,b]{Valentina Viotto}
\author[a,b]{Roberto Ragazzoni}
\author[a,b]{Carmelo Arcidiacono}
\author[a,b]{Maria Bergomi}
\author[a,b]{Davide Greggio}
\author[a,b]{Kalyan Radhakrishnan}
\author[a,b]{Simone di Filippo}
\author[a,b]{Luca Marafatto}
\author[a,b]{Marco Dima}
\author[a,b]{Federico Biondi}
\author[a,b]{Jacopo Farinato}
\author[a,b]{Demetrio Magrin}

\affil[a]{INAF - Osservatorio Astronomico di Padova - vicolo dell'Osservatorio, 35122, Padova, Italy}
\affil[b]{ADONI - Laboratorio Nazionale di Ottica Adattiva, Italy}

\authorinfo{Further author information: (Send correspondence to EP)\\EP: E-mail: elisa.portaluri@inaf.it}

\begin{document} 
\maketitle

\begin{abstract}
  The Ingot WFS, a Pyramid-like WFS, has been proposed as a relief to the LGS spot elongation.
  Actually, the artificial sources are confined in an excited cigar-shaped region in the sodium layer and portions of the telescope aperture have a corresponding different perspective. This diversity generates a variation of the WFS response depending on the illuminated sub-aperture position. The feasibility study of the INGOT WFS is developing within the MAORY project.
In this work we present the numerical simulator built in order to investigate the performance of the Ingot WFS in terms of Strehl Ratio, obtained reconstructing the incoming turbulent wavefront with a modal approach, in a closed-loop fashion.
We also discuss the assumptions and tests we made in order to explore the range of parameters that play key roles in the game. Finally, we report on the overall results of the simulations of the performance expected by the Ingot WFS, under different conditions and input aberrations, with the aim to also compare the measured AO loop residuals with the Shack-Hartmann WFS ones, obtained working under the same assumptions.
\end{abstract}


\section{INTRODUCTION}
\label{sec:intro} 
In 2017, Ragazzoni et al. [\citenum{Ragazzoni2017}] presented a new idea of wavefront sensor, literally ``a new fashion for an old paradigm". Such device was thought to overcome some limitations that affect the adaptive optics facilities that foresee the use of laser guide stars (LGSs) as references for the wavefront sensing.
There we started from a 6-faces configurations, making the object very similar to an ingot, and hence the name came.

This kinf of device is intended to be used on the next generation of extremely large telescopes MCAO modules, which, of course, are going to be fed by artificial sources, like the European Extremely Large Telescope [\citenum{Gilmozzi2007}], the Giant Magellan Telescope [\citenum{Johns2008}], and the Thirty Meter Telescope [\citenum{Szeto2008}].
LGSs, already used in modern telescopes, show some differences from natural stars, first of all being not point-like objects, but like 'cigars' in the sky because they are generated by the excitation of the Sodium layer, present in the atmosphere at around 90 km, therefore at a finite distance and with a certain thickness.
Moreover this cigar has not the same shape as seen from each sub-aperture of a 40-m telescope: its geometry varies depending on the angle under which we see the LGS and gives rise to some important effects like the spot elongation (and the need for truncation if we consider a Shack-Hartmann wavefront sensor, SH-WFS, [\citenum{Beckers1989}]), which play an important role when measuring the total noise in order to reconstruct and correct the wavefront.
We should add to these considerations the fact that the cigar would not focus on a plane, but on a 3D volume! As the geometry is very important when we consider the total error budget of the system, the Ingot wavefront sensor (I-WFS) takes into account this problem.
These issues (and other ones, like the fact that the source is monochromatic) are well known and massively studied in literature (i.e.:[\citenum{Fried1995},\citenum{Pfrommer2009},\citenum{Diolaiti2012}]).

\begin{figure} [h]
   \begin{center}
   \begin{tabular}{c} 
   \includegraphics[height=4cm]{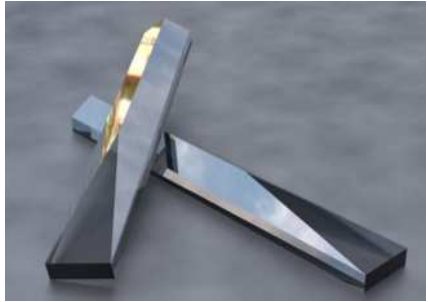}
   \end{tabular}
   \end{center}
   \caption[example] 
   { \label{fig:ingot} 3-faces Ingot prism design.}
\end{figure}

In the meanwhile the ingot became a 3-faces sensor (Figure~\ref{fig:ingot}) with 2 reflecting and 1 transmitting parts (see [\citenum{Ragazzoni2018},\citenum{Ragazzoni2019}] for an update of the sensor characteristics and idea).
To evaluate the impact of this innovative WFS compared with a classical SH-WFS, we set up a group of people taking care of the different aspects necessary to a feasibility study, i.e. arranging first laboratory experiments, fine-tuning the geometrical characteristics, working on the optical design and starting first simulations of the performance.
In this work I will focus on the last aspect, describing the tomographic simulator we set up to study a 40-m telescope like the ELT equipped with an I-WFS and the first results we obtained.

\section{The simulation tool}
We developed a simulation tool [\citenum{Viotto2018}], partially following a common approach, analogous to the one used by other tools dedicated to other WFSs, but different in some aspects because of the peculiar shape of the LGSs and of the geometrical characteristics of the I-WFS itself. Some of these differences are discussed in [\citenum{Viotto2019}], together with the description of the trade-off we made to chose between the Fourier transform approach and the ray tracing one.

\begin{figure} [h]
   \begin{center}
   \begin{tabular}{c} 
   \includegraphics[height=7.5cm]{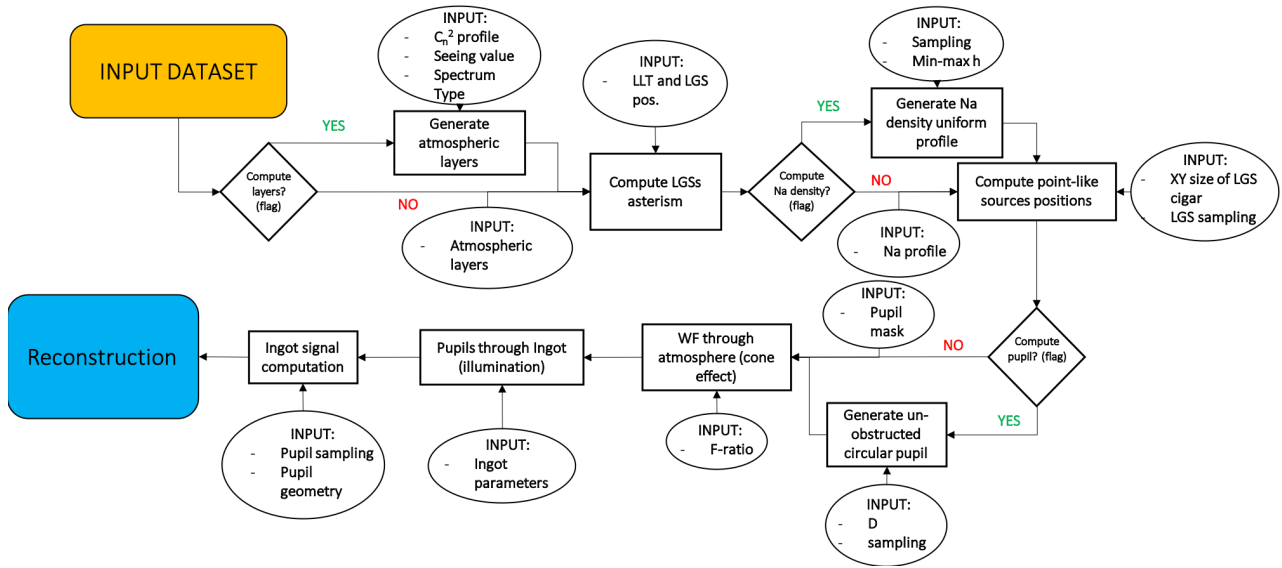}
   \end{tabular}
   \end{center}
   \caption[example] 
   { \label{fig:diag} Logical flow of the simulation tool, including inputs and flags.}
\end{figure}

The code is written in Interactive Data Language (IDL) version 8.7 and it is composed by a number of functions computing different steps required by the overall simulation. The block diagram in Figure~\ref{fig:diag} describes the logical flow of the information inside the code, including inputs to be used at different stages and different flags that can be turned on depending on the aim of the simulation.

\begin{figure} [h]
   \begin{center}
   \begin{tabular}{c} 
   \includegraphics[height=6cm]{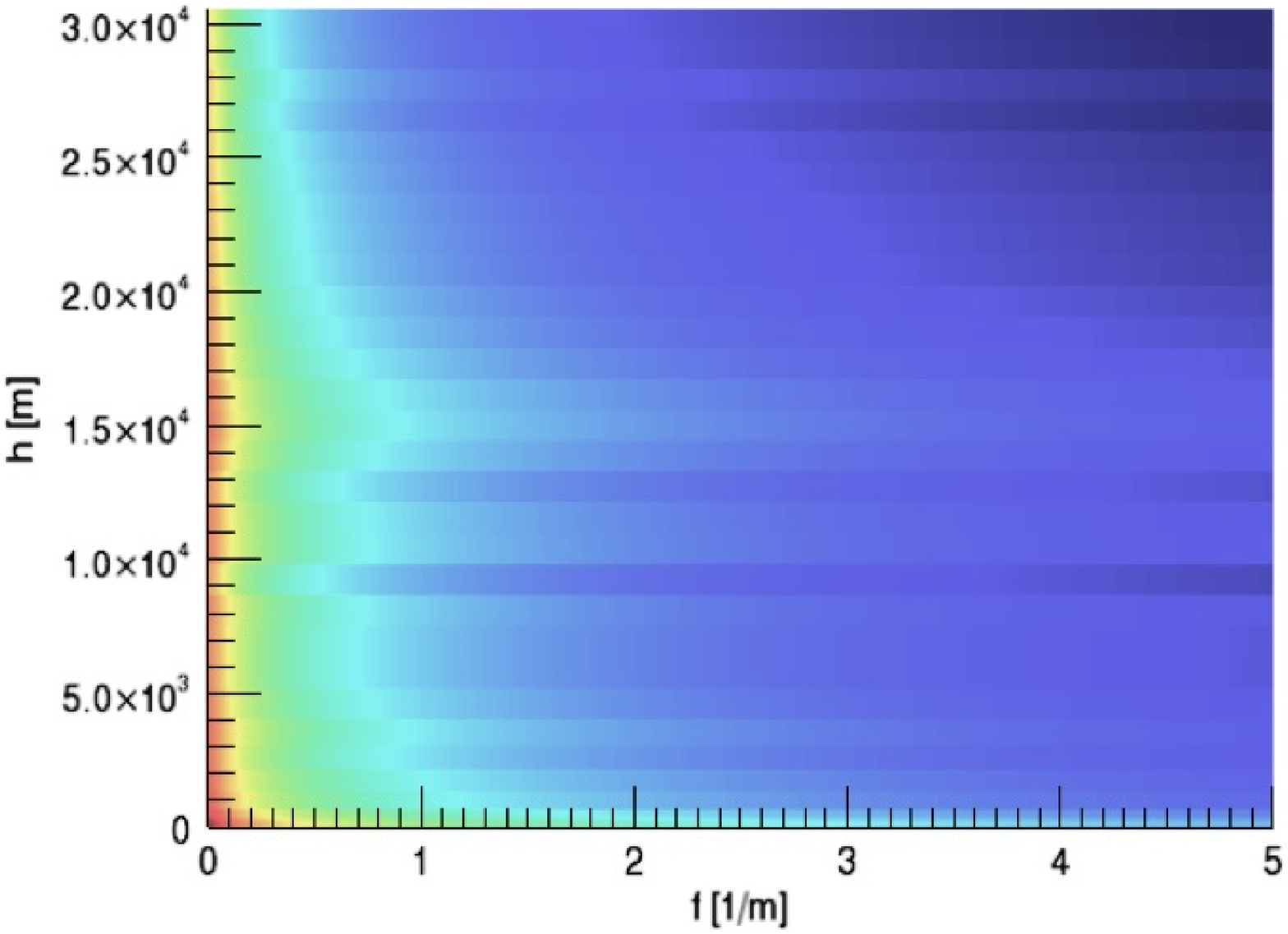}  \hspace{2cm} \includegraphics[height=5cm]{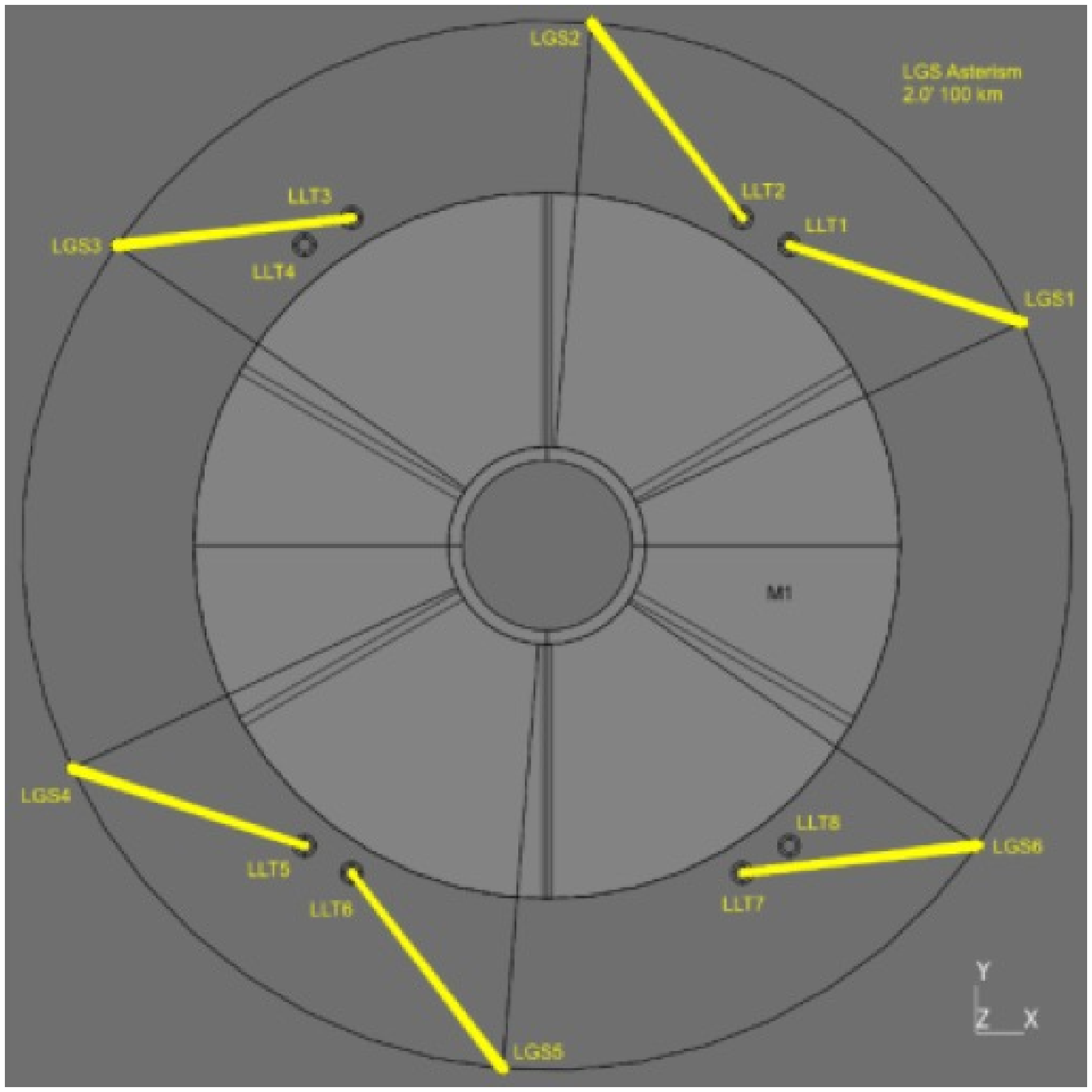}
   \end{tabular}
   \end{center}
   \caption[example] 
   { \label{fig:in1} Left-hand panel: $C_n^2$ profile at Paranal adopted by ESO. It is composed by 35 turbulence layers. Right-hand panel: geometrical configuration of the launchers that fires the lasers.}
\end{figure}

\begin{figure} [h]
   \begin{center}
   \begin{tabular}{c} 
   \includegraphics[height=6cm]{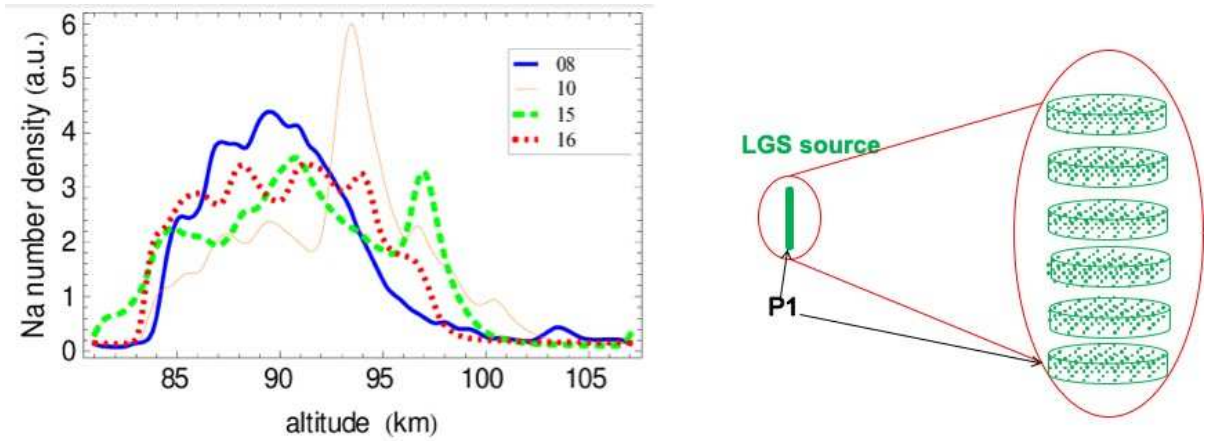}
   \end{tabular}
   \end{center}
   \caption[example] 
   { \label{fig:in2} Left-hand panel: several Sodium profiles that describe the thickness of the layer. Right-hand panel: an example of the LGS sampling.}
\end{figure}

The simulation starts building an input wavefront that is perturbed and ends reconstructing it and calculating the residuals between input and output:
\begin{equation}
  WF_{residual}=\frac{WF_{input}-WF_{reconstructed}}{WF_{input}}.
  \end{equation}
The perturbation is driven by a certain atmosphere we use (the left-hand panel of Figure~\ref{fig:in1} shows for example the 35-layers $C_n^2$ profile at Paranal) and some other tomographic inputs like the Fried parameter $r_0$ and the power spectrum selected (von Karman or Kolmogorov). From such information the code generates the layers and takes also into account the telescope and launcher characteristics like size and position (see right-hand panel of Figure~\ref{fig:in1}) to understand the asterism considered.
For the moment we are considering only 1 LGS with a given sodium profile and with a certain spatial sampling (Figure~\ref{fig:in2}): we can select between random sample to fill the source, or representing it like a cylinder uniformly sampled or with a number of disks with a certain intensity profile (this option is shown in the figure and it is used for the most of the time). The details of how we choose such a sampling are extensively described in [\citenum{Viotto2019}], as a good compromise between speed of the code and resolution of the simulation.

\begin{figure} [h]
   \begin{center}
   \begin{tabular}{c} 
   \includegraphics[height=6cm]{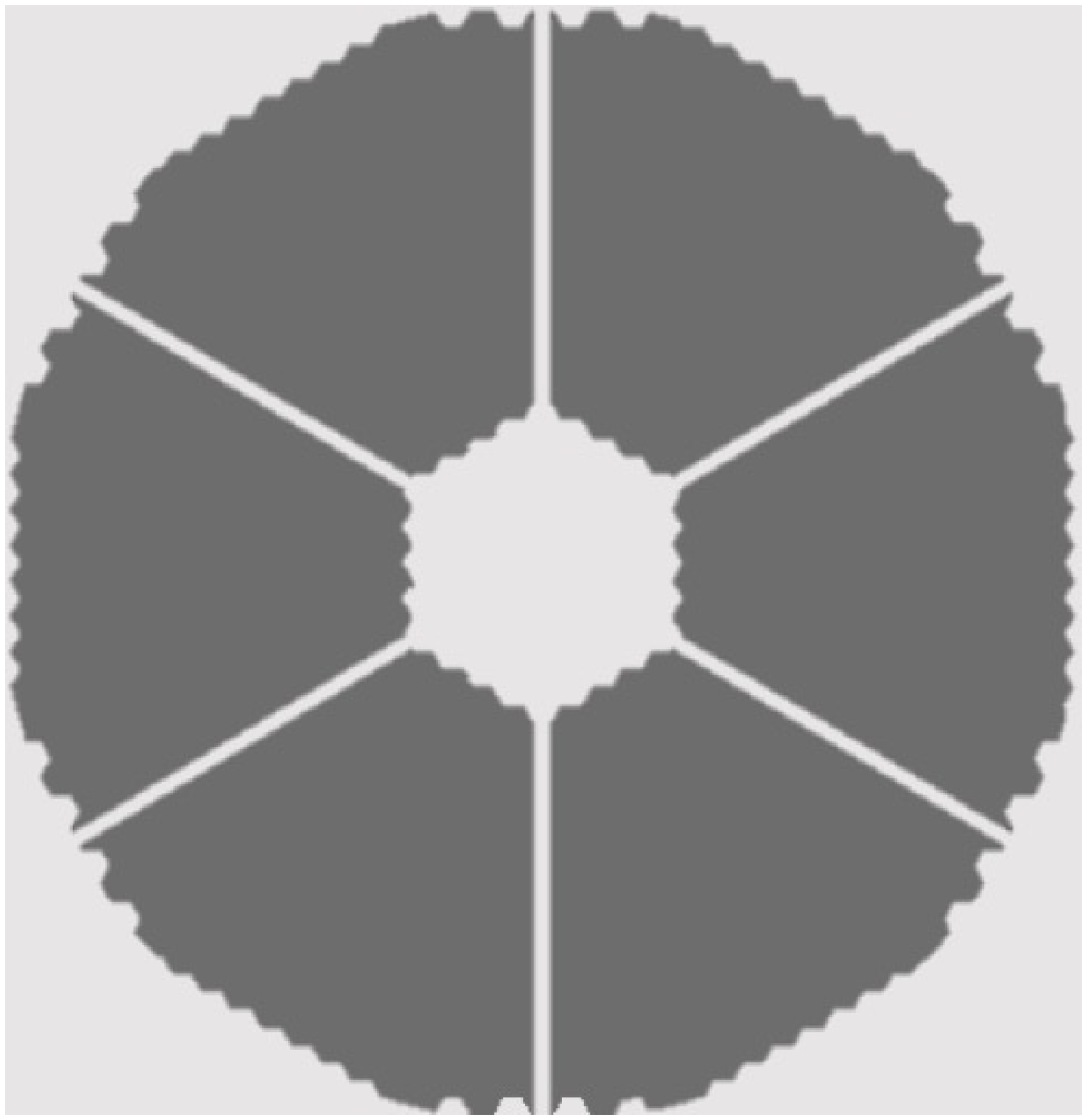}
   \includegraphics[height=6cm]{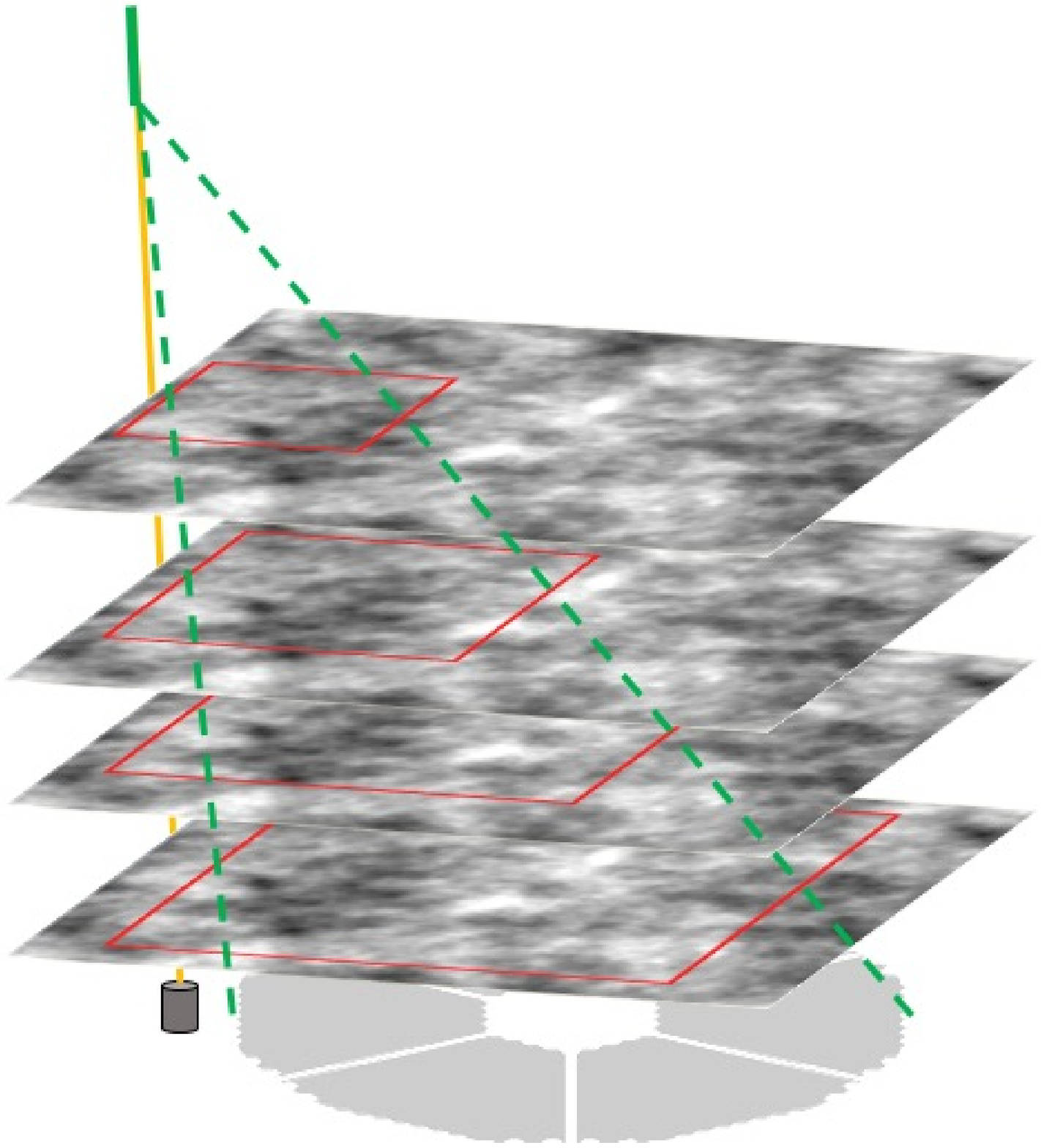}
   \includegraphics[height=6cm]{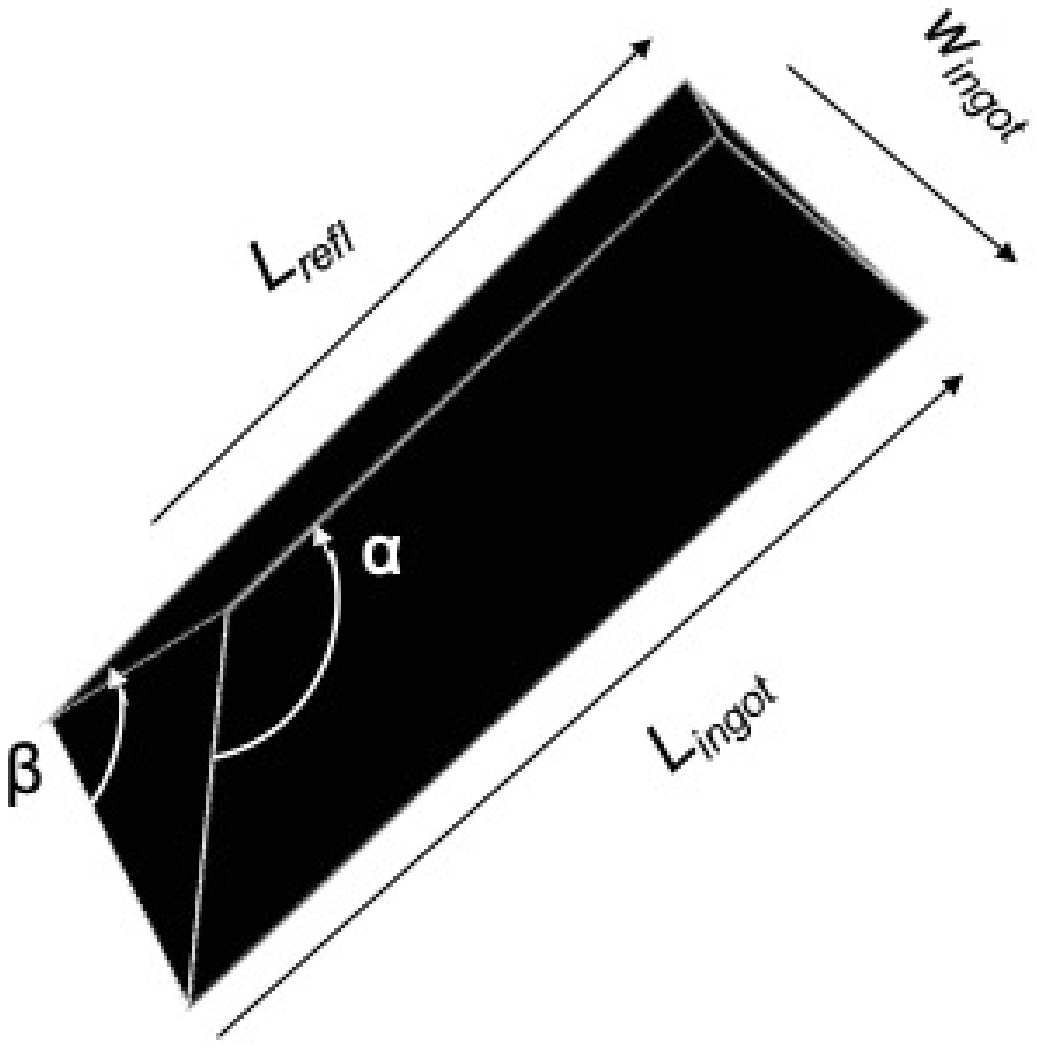}
   \end{tabular}
   \end{center}
   \caption[example] 
   { \label{fig:in3}Left-hand panel: pupil mask of the ELT used in the simulation. Central panel: representation of the cone effect due to the finite distance of the source. The light of the laser intercepts only a portion of the different atmospheric layers. Right-hand: geometrical parameters of the I-WFS describing the size of the sensor and of the reflecting parts.}
\end{figure}

Then the code proceeds characterizing the system with the resolution required, the desired binning, the apertures mask, and other information needed, also considering the cone effect due to the finite distance of the source to select the portion of the atmosphere to be considered (Figure~\ref{fig:in3}).

The final step is considering the geometrical characteristics of the sensor, i.e. the I-WFS (Figure~\ref{fig:in3}) and, adding all the information, the code measures how the pupils are illuminated by each sub-aperture and calculates the signal as follows:
\begin{equation}
  S_x=\frac{A-B}{A+B}, \hspace{2cm}
  S_y=\frac{A+B-C}{A+B+C}
\end{equation}
where A and B are the reflective pupils, while C is the transmitted one.

These signals are reconstructed using the interaction matrix, built using a Zernike basis through the same simulator with consistent parameters. For all the tests and runs we used the telescope diameter $D = 38.5$ m and $\lambda = 0.589$~$\mu$m.

\section{Tests and first results}
\begin{figure} [h]
   \begin{center}
   \begin{tabular}{c} 
    \includegraphics[width=17.5cm]{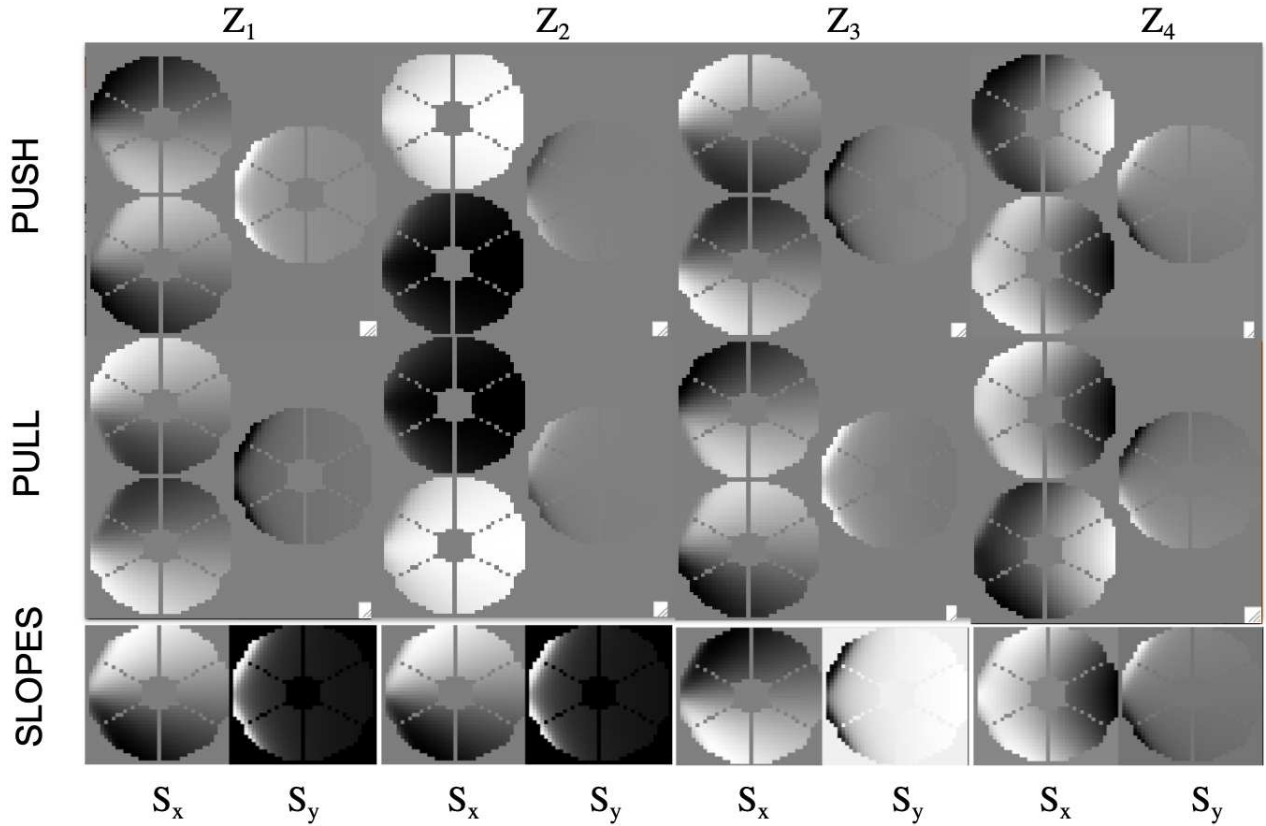}\\
   \end{tabular}
   \end{center}
   \caption[example] 
   { \label{fig:rec} Top panel: 3 ingot pupils corresponding to the first 4 Zernike modes: on the first line we have the push movement, while second line represents the pull one. Bottom panel: 2 ingot signals measured from the 3 ingot pupils corresponding to the first 4 Zernike modes: on the left we have $S_X$, while $S_y$ is represented on the right.}
\end{figure}

In this section we describe the first tests we performed with the simulator. We used a Zernike basis with 100 modes: Figure~\ref{fig:rec} shows as an example how the 3 pupils and the 2 signals look like for the first 4 modes.

\begin{figure} [h]
   \begin{center}
   \begin{tabular}{c} 
   \includegraphics[height=6cm]{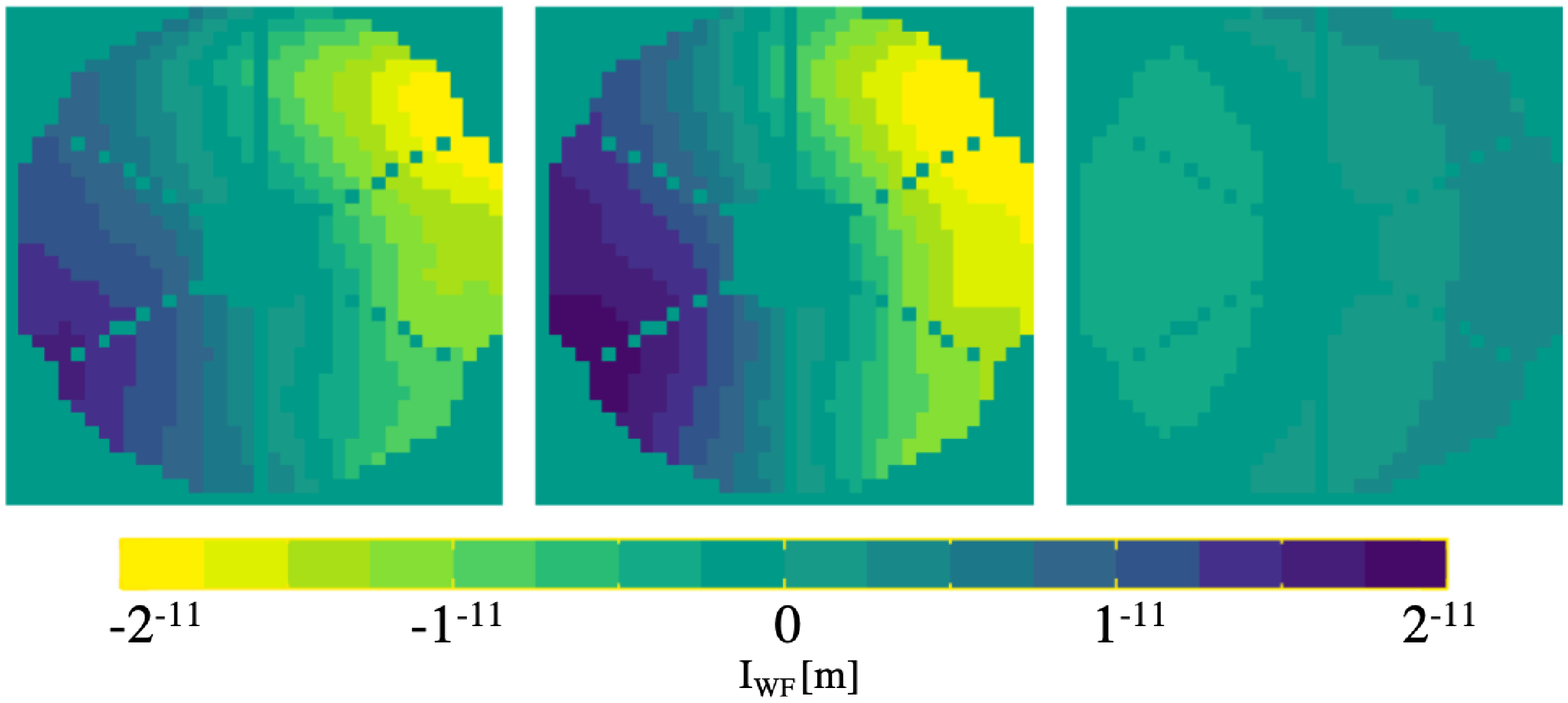}\\
   \includegraphics[height=8cm]{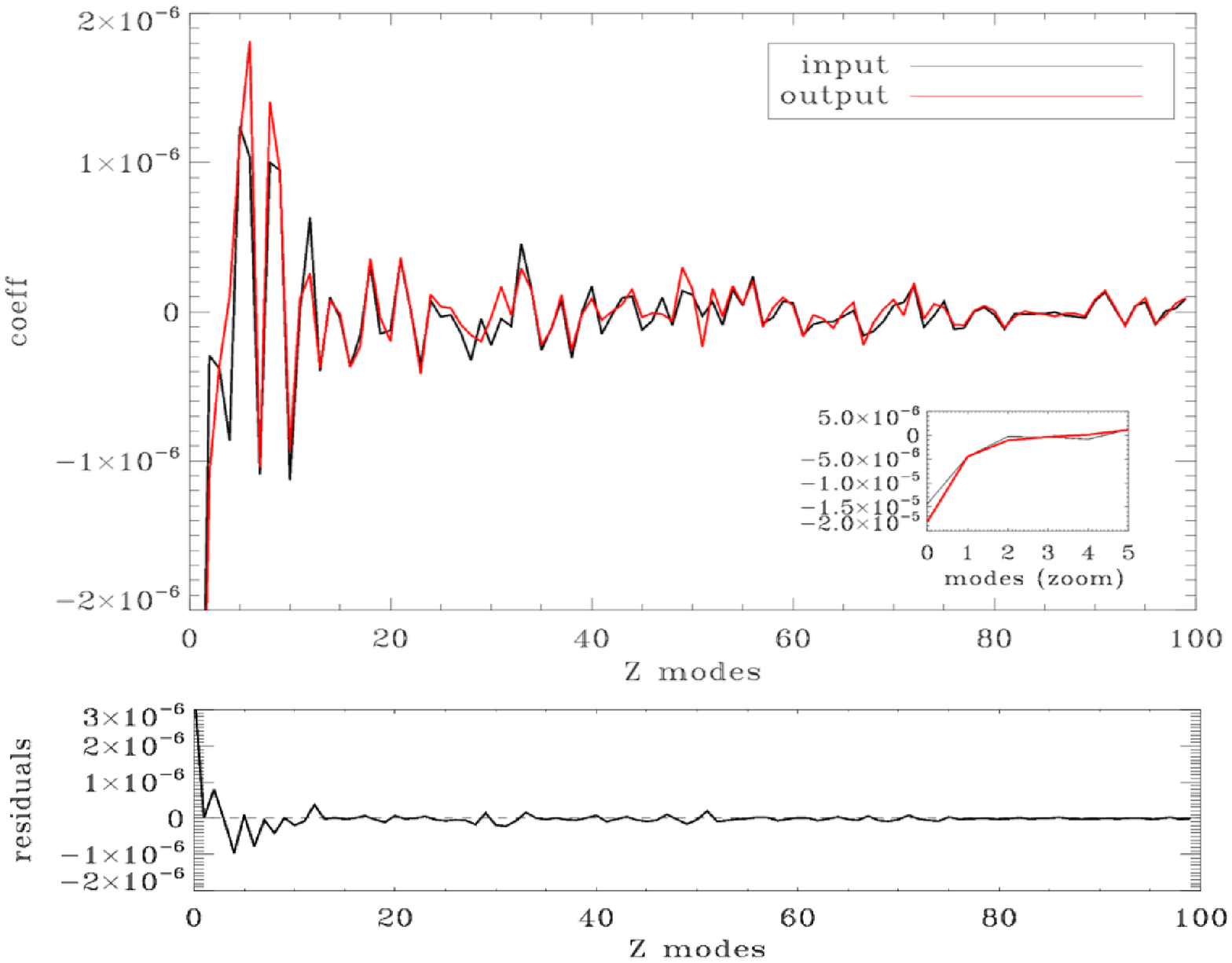}
    \end{tabular}
   \end{center}
   \caption[example] 
   { \label{fig:oloop} Example of a reconstruction. Top panel: the input wavefront (on the left) is built linearly combining the 100 Zernike modes basis using Noll coefficients and it is reconstructed by the simulator (central panel). On the left the residuals between input and output are shown using a different scale to highlight the features inside. Bottom panel: 100 Noll coefficients used as input (black line) and recovered after the reconstruction (red line) in one rune of the code, so in open loop. The bottom plot represents the difference between them.}
\end{figure}

Once we have built the basis, we can do a first reconstruction (Figure~\ref{fig:oloop}): the input wavefront was generated linearly combining the 100 Zernike modes with coefficients taken considering the Noll sequential indices [\citenum{Noll1976}], as also shown in the plot of the same Figure.
The residuals between input and output wavefronts are low, of the order of 20-30\%, therefore the results are very promising, especially if we consider that they are obtained after only one run of the reconstruction, so in an open loop fashion.

\begin{figure} [h]
   \begin{center}
   \begin{tabular}{c} 
   \includegraphics[width=17cm]{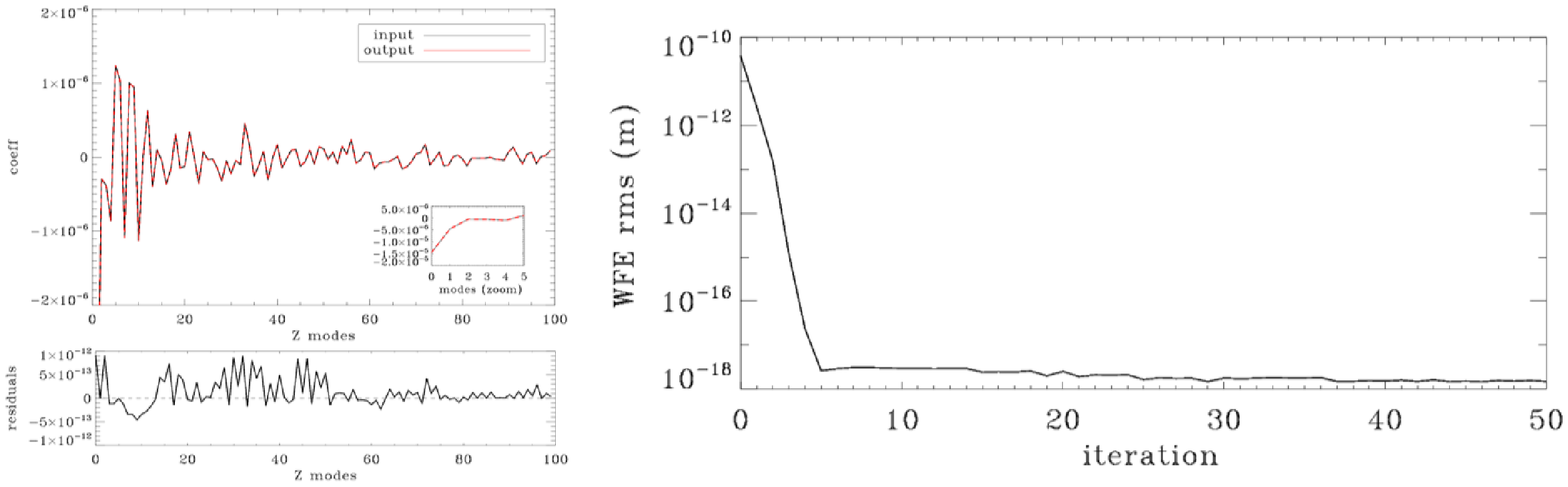}
    \end{tabular}
   \end{center}
   \caption[example] 
   { \label{fig:cloop} Example of a reconstruction in a close loop. Left-hand panel: 100 Noll coefficients used as input (black line) and recovered after closing the loop (red line). The bottom plot represents the difference between them. Right-hand panel: wavefront error measured as the rms of the residuals as a function of the iterations.}
\end{figure}

Next step is try to close a proper loop to see whether it converges and evaluate the performance.
Figure~\ref{fig:cloop} shows that case using the same input wavefront described above and running 50 times the code. Looking at the wavefront error, measured from the residuals rms, we can consider the loop closed even before the 50 iterations, at around 6, and the Noll coefficients are perfectly recovered. 
However, it is important to mention that these tests were made using a ``frozen" atmosphere, i.e. without dynamic disturbance, so in the infinite computing power (i.e. extremely optimistic and less realistic) case.

\begin{figure} [h]
   \begin{center}
   \begin{tabular}{c} 
   \includegraphics[height=4.2cm]{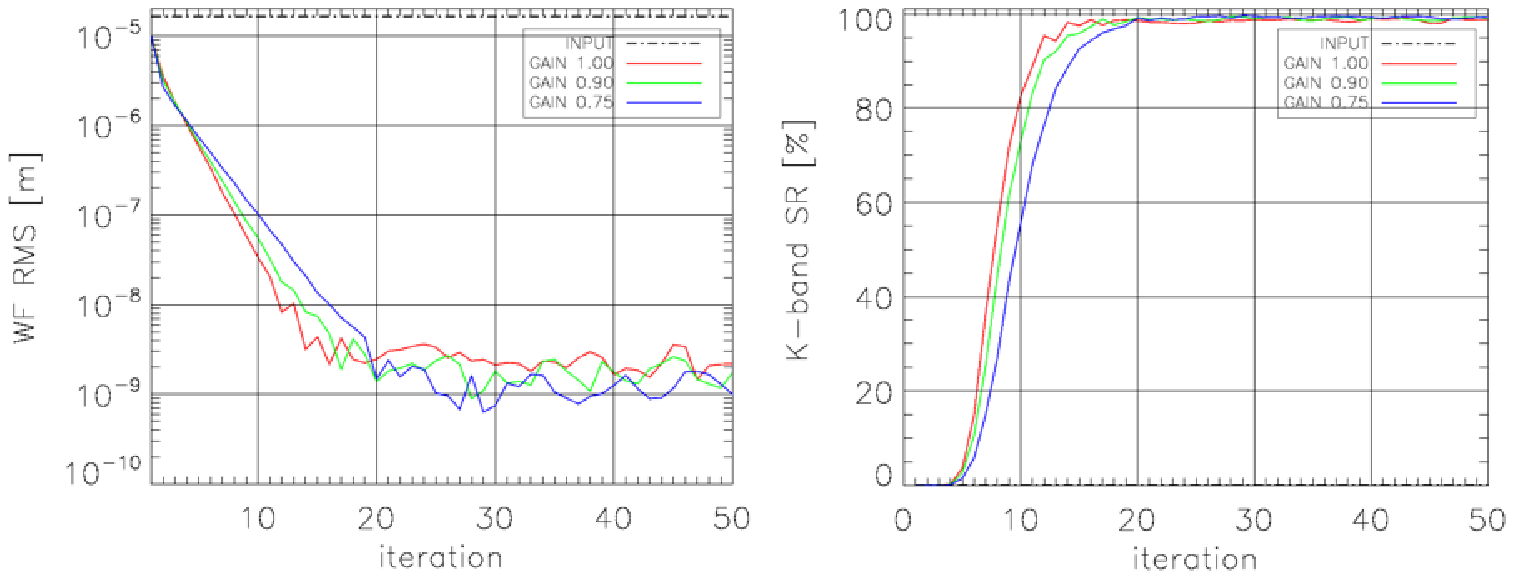}\hspace{1cm}
   \includegraphics[height=4.2cm]{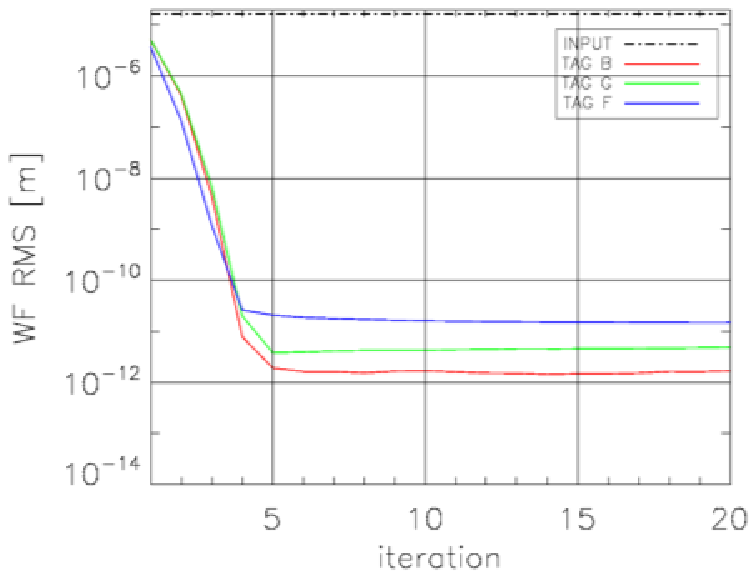}\\
   \includegraphics[width=8.5cm]{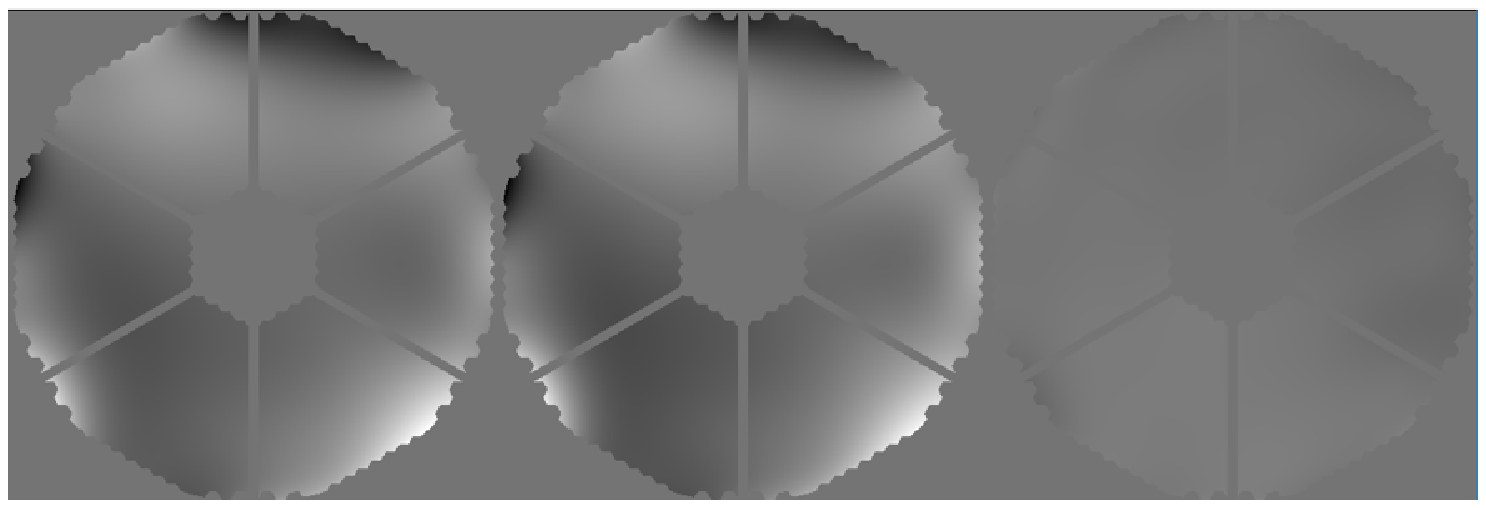}
   \includegraphics[width=8.5cm]{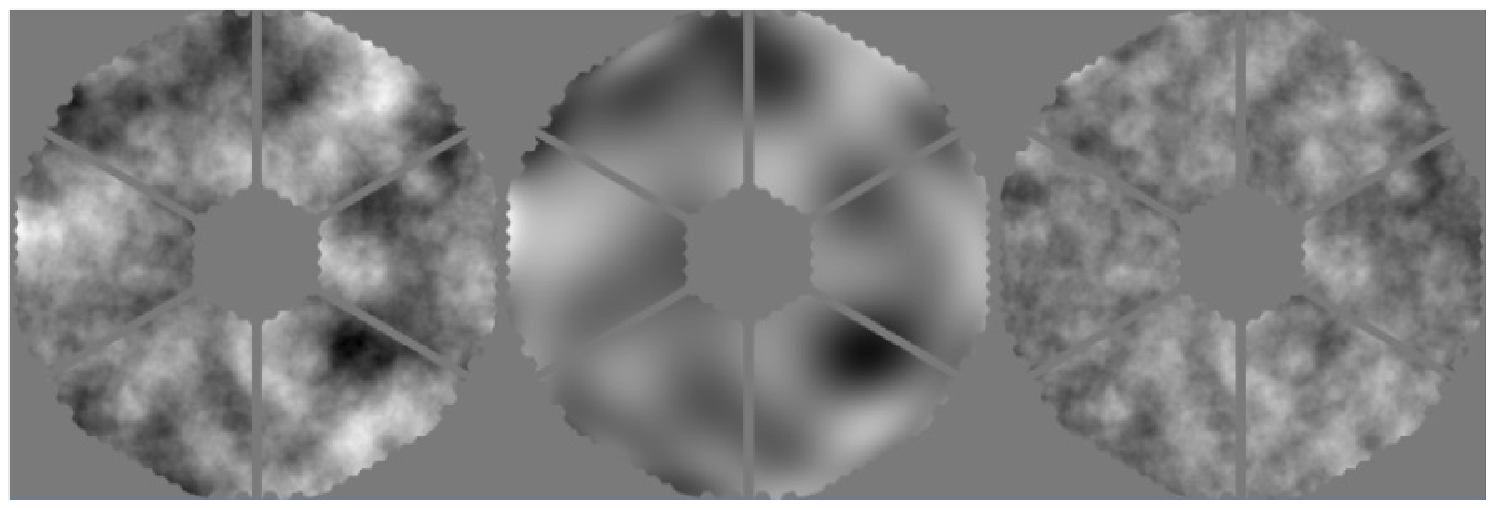}
   \end{tabular}
   \end{center}
   \caption[example] 
   { \label{fig:test} Top panel, from left to right: wavefront error rms and corresponding K-band Strehl ratio measured for different gains used for the reconstruction, and wavefront error rms obtained changing the sampling of the source. Bottom panel: reconstruction run with input, output, and residual wavefront obtained with a different resolution with respect to Figure~\ref{fig:cloop} (on the left), and using a full turbulence as input (on the right). The greyscale is kept the same to enanche the internal features.}
\end{figure}

But once we tested that in the basic condition the code is running (and the I-WFS is working properly), we can vary some parameters to see how the results change and how stable/fast is the code.
Figure~\ref{fig:test} shows first results obtained changing the sampling of the source, the gain and the resolution, plus a first attempt of reconstruction using a turbulent wavefront, i.e. built from the given atmosphere and not as a combination of Zernike modes.

After a period of debugging and investigation of some issues, now all the outcome pointed in the right direction. At the moment we are working to increase the speed of the code in order to make more tests and evaluation before starting a proper end-to-end simulation, which we aim to compare with one obtained using a SH wavefront sensor to show how a 40-m telescope can benefit using the I-WFS.

\section{Conclusions}

We have set up a group working on an innovative wavefront pupil plane sensor for extended 3D sources and we are progressing in different aspects of its feasibility study:
\begin{itemize}
\item We have built an end-to-end simulator to test the performance of the I-WFS in the ELT configuration. It is stable and works properly.
\item We have demonstrated that we were able to close the loop and perform a good reconstruction of the wavefront at least for a static turbulence.
\item We are improving the simulations changing the input parameters as resolution, sampling, number of modes and dynamical disturbance to have a full description of the performance. This is going parallel with the need of speed-up the code.
\item We will compare the results with those obtained using SH-WFS and with those obtained in the laboratory (first results are described in \citenum{diFilippo2019}) to evaluate properly the positive impact of a I-WFS on a 40-m telescope.
\end{itemize}

\bibliography{main} 

\begin{thebibliography}{10}

\bibitem{Ragazzoni2017}
{Ragazzoni}, R., {Portaluri}, E., {Viotto}, V., {Dima}, M., {Bergomi}, M.,
  {Biondi}, F., {Farinato}, J., {Carolo}, E., {Chinellato}, S., {Greggio}, D.,
  {Gullieuszik}, M., {Magrin}, D., {Marafatto}, L., and {Vassallo}, D.,
  ``{Ingot Laser Guide Stars Wavefront Sensing},'' {\em AO4ELT5} ,
  arXiv:1808.03685 (Aug 2017).

\bibitem{Gilmozzi2007}
{Gilmozzi}, R. and {Spyromilio}, J., ``{The European Extremely Large Telescope
  (E-ELT)},'' {\em The Messenger}~{\bf 127} (Mar. 2007).

\bibitem{Johns2008}
{Johns}, M., ``{The Giant Magellan Telescope (GMT)},'' in [{\em Extremely Large
  Telescopes: Which Wavelengths? Retirement Symposium for Arne
  Ardeberg}{\nolinebreak\hspace{0.1em}]},  {\em SPIE} {\bf 6986},  698603 (Apr.
  2008).

\bibitem{Szeto2008}
{Szeto}, K., {Roberts}, S., {Gedig}, M., {Austin}, G., {Lagally}, C.,
  {Patrick}, S., {Tsang}, D., {MacMynowski}, and et~al., ``{TMT telescope
  structure system: design and development progress report},'' in [{\em
  Ground-based and Airborne Telescopes II}{\nolinebreak\hspace{0.1em}]},  {\em
  SPIE} {\bf 7012},  70122G (July 2008).

\bibitem{Beckers1989}
{Beckers}, J.~M., ``{Detailed compensation of atmospheric seeing using
  multiconjugate adaptive optics.},'' in [{\em
  SPIE}{\nolinebreak\hspace{0.1em}]},  {Roddier}, F.~J., ed., {\em Society of
  Photo-Optical Instrumentation Engineers (SPIE) Conference Series} {\bf 1114},
   215--217 (Sep 1989).

\bibitem{Fried1995}
{Fried}, D.~L., ``{Focus anisoplanatism in the limit of infinitely many
  artificial-guide-star reference spots.},'' {\em Journal of the Optical
  Society of America A}~{\bf 12} (May 1995).

\bibitem{Pfrommer2009}
{Pfrommer}, T., {Hickson}, P., and {She}, C.-Y., ``{A large-aperture sodium
  fluorescence lidar with very high resolution for mesopause dynamics and
  adaptive optics studies},'' {\em Geophysical Research Letters}~{\bf 36},
  L15831 (Aug. 2009).

\bibitem{Diolaiti2012}
{Diolaiti}, E., {Schreiber}, L., {Foppiani}, I., and {Lombini}, M.,
  ``{Dual-channel multiple natural guide star wavefront sensor for the E-ELT
  multiconjugate adaptive optics module},'' in [{\em Adaptive Optics Systems
  III}{\nolinebreak\hspace{0.1em}]},  {\em SPIE} {\bf 8447},  84471K (July
  2012).

\bibitem{Ragazzoni2018}
{Ragazzoni}, R., {Greggio}, D., {Viotto}, V., {Di Filippo}, S., {Dima}, M.,
  {Farinato}, J., {Bergomi}, M., {Portaluri}, E., {Magrin}, D., {Marafatto},
  L., {Biondi}, F., {Carolo}, E., {Chinellato}, S., {Umbriaco}, G., and
  {Vassallo}, D., ``{Extending the pyramid WFS to LGSs: the INGOT WFS},'' in
  [{\em SPIE}{\nolinebreak\hspace{0.1em}]},  {\em Society of Photo-Optical
  Instrumentation Engineers (SPIE) Conference Series} {\bf 10703},  107033Y
  (Jul 2018).

\bibitem{Ragazzoni2019}
{Ragazzoni}, R., ``{Pupil plane wavefront sensing for extended and 3D
  sources},'' {\em AO4ELT6}  (2019).

\bibitem{Viotto2018}
{Viotto}, V., {Portaluri}, E., {Arcidiacono}, C., {Ragazzoni}, R., {Bergomi},
  M., {Di Filippo}, S., {Dima}, M., {Farinato}, J., {Greggio}, D., {Magrin},
  D., and {Marafatto}, L., ``{Dealing with the cigar: preliminary performance
  estimation of an INGOT WFS},'' in [{\em SPIE}{\nolinebreak\hspace{0.1em}]},
  {\em Society of Photo-Optical Instrumentation Engineers (SPIE) Conference
  Series} {\bf 10703},  107030V (Jul 2018).

\bibitem{Viotto2019}
{Viotto}, V., {Portaluri}, E., {Arcidiacono}, C., {Bergomi}, M., {Di Filippo},
  S., {Dima}, M., {Greggio}, D., {Farinato}, J., {Magrin}, D., Kalyan, R., and
  {Ragazzoni}, R., ``{INGOT Wavefront Sensor: simulation of pupil images},''
  {\em AO4ELT6}  (2019).

\bibitem{Noll1976}
{Noll}, R.~J., ``{Zernike polynomials and atmospheric turbulence.},'' {\em
  Journal of the Optical Society of America (1917-1983)}~{\bf 66},  207--211
  (Mar 1976).

\bibitem{diFilippo2019}
{Di Filippo}, S., {Greggio}, D., {Bergomi}, M., {Radhakrishnan}, K.,
  {Portaluri}, E., {Viotto}, V., {Arcidiacono}, C., {Magrin}, D., {Ragazzoni},
  R., {Janin-Potiron}, P., {Schatz}, L., {Neichel}, B., and {Fusco}, t.,
  ``{INGOT Wavefront Sensor: from the optical design to a preliminary
  laboratory test},'' {\em AO4ELT6}  (2019).

\end{thebibliography}
\bibliographystyle{spiebib} 

\end{document}